# Current-driven Dynamics of Skyrmion Bubbles in Achiral Uniaxial Magnets

Yaodong Wu(吴耀东)[1,3], Jialiang Jiang(蒋佳良)[3], and Jin Tang(汤进)[2,3*]

[1]School of Physics and Materials Engineering, Hefei Normal University, Hefei 230601, China

[2]School of Physics and Optoelectronics Engineering Science, Anhui University, Hefei, 230601, China

[3]Anhui Province Key Laboratory of Condensed Matter Physics at Extreme Conditions, High Magnetic Field Laboratory, HFIPS, Anhui, Chinese Academy of Sciences, Hefei, 230031, China

*Corresponding author: jintang@ahu.edu.cn

**Abstract**

We report dynamics of skyrmion bubbles driven by spin-transfer torque in achiral ferromagnetic nanostripes using micromagnetic simulations. In a three-dimensional uniaxial ferromagnet with a quality factor that is smaller than 1, the skyrmion bubble is forced to stay at the central nanostripe by a repulsive force from the geometry border. The coherent motion of skyrmion bubbles in the nanostripe can be realized by increasing the quality factor to ~3.8. Our results should propel the design for future spintronic devices such as artificial neural computing and racetrack memory based on dipole-stabilized skyrmion bubbles.

## 1. Introduction

Skyrmions are promising information carriers applied in future high-performance spintronic devices owing to their particle-like and topological properties.[1-9] Particularly, the easy dynamic manipulations of skyrmions at very small currents predict the potential energy-efficient functional operations in skyrmion-based spintronic devices such as racetrack memory.[10-12] Skyrmions are typically stabilized by chiral Dzyaloshinsky-Moriya interactions in non-centrosymmetric magnets.[1,4] The spin-torque-driven motion dynamics of chiral skyrmions such as velocity and Hall angle have been well constructed in theory and experimentally identified.[13-17] Besides, skyrmions (also called skyrmion bubbles) can be also stabilized by dipole-dipole interactions in centrosymmetric uniaxial magnets.[2,18-25] There are two types of bubbles in centrosymmetric uniaxial magnets, i.e. type-I skyrmion bubble and type-II topologically-trivial bubble.[24] Because of the absence of chiral interactions, skyrmion bubbles in the centrosymmetric magnets are achiral and have two helictites at a given field.[19,20] Achiral skyrmion bubbles have been found in many room-temperature centrosymmetric ferromagnets, such as $Fe_3Sn_2$[22,26], Fe/Gd[20], and BaFeScMgO.[21] The size of achiral skyrmion bubbles can be as small as ~50 nm and is comparable with room-temperature chiral skyrmions.[20,27] In previous bubble memory, the motion of achiral skyrmion bubbles can be driven by Oster fields,[23,28,29] which, however, is not superior in now spintronic devices. Recent studies have shown current-induced emergent dynamic manipulations of achiral skyrmion bubbles, such as topological skyrmion-bubble transformations.[19,30] However, the current-induced coherent dynamic motions and their physical mechanisms of achiral skyrmion bubbles are less demonstrated.

Here, using a micromagnetic simulation approach,[31] we present the dynamic behaviors of achiral skyrmion bubbles driven by Zhang-Li spin-transfer-torque in uniaxial ferromagnetic stripes.[32] The role of dipole interaction in preventing the skyrmion from the central nanostripe and its potential application for neural computing is demonstrated.[33-36] Our results provide further insights into the current-induced dynamic manipulation of achiral dipole-stabilized skyrmion bubbles.

## 2. Methods

We used MuMax3 to simulate the current-induced dynamics of achiral skyrmion bubbles.[31] The total energy terms are given by:

$$\varepsilon = \int_{V_s} \{\varepsilon_{\text{ex}} + \varepsilon_{\text{u}} + \varepsilon_{\text{zeeman}} + \varepsilon_{\text{dem}}\} d\boldsymbol{r} \tag{1}$$

Here, exchange energy $\varepsilon_{\text{ex}} = A(\partial_x \mathbf{m}^2 + \partial_y \mathbf{m}^2 + \partial_z \mathbf{m}^2)$, uniaxial magnetic anisotropy energy $\varepsilon_{\text{u}} = -K_{\text{u}}(\mathbf{m} \cdot \mathbf{n}_{\text{u}})^2$, Zeeman energy $\varepsilon_{\text{zeeman}} = -M_s \mathbf{B}_{\text{ext}} \cdot \mathbf{m}$, and demagnetization energy $\varepsilon_{\text{dem}} = -\frac{1}{2} M_s \mathbf{B}_{\text{d}} \cdot \mathbf{m}$. Here, $\mathbf{m} \equiv \mathbf{m}(x, y, z)$ is the normalized units continuous vector field that represents the magnetization $\mathbf{M} \equiv M_s \mathbf{m}(x, y, z)$. The parameters $A$, $K_{\text{u}}$, and $M_s$ are the exchange interaction, uniaxial anisotropy constant, and saturation magnetization, respectively. $\mathbf{n}_{\text{u}}$ is the unit vector field of the uniaxial easy magnetization axis and is set along the (001) $z$-axis. $\mathbf{B}_{\text{d}}$ is the demagnetization field. The magnetic parameters were set based on typical bubble-hosting materials with $M_s$ = 622.7 kA/m, exchange constant $A$ = 8.25 pJ/m.[18] The length, width, and height of nanostripes are set 2000 nm, 500 nm, and 100 nm, respectively. We set the cell size to 2 × 2 × 2 nm³. We modify only $K_u$ to discuss the role of dipole-dipole interaction and the quality factor. The Zhang-Li spin-transfer-torque is considered to drive the motion of skyrmion bubbles[32]. For simplicity, we exclude the skyrmion Hall effect in our simulation by assuming that the Gilbert damping factor $\alpha$ is equal to non-adiabatic parameter $\beta$, i.e. $\alpha = \beta = 0.1$. The location of the skyrmion is obtained by calculating the guiding center $G_x$ along the $x$ axis, which is defined by the moments of the topological charge density as[26]: $G_x = \frac{\iint xq \mathrm{d}x\mathrm{d}y}{\iint q\mathrm{d}x\mathrm{d}y}$. Here, $q$ is the topological charge density expressed as: $q = \frac{1}{2}\varepsilon_{xy}(\partial_x \mathbf{m} \times \partial_y \mathbf{m})\cdot \mathbf{m}$, where $\varepsilon_{xy}$ is the antisymmetric tensor.

3. **Results and discussions**

Figure 1 shows the current-driven dynamic motions of a skyrmion in the ferromagnetic nanostripe with $K_u$ = 54.5 kJ m⁻³. The skyrmion is initialed at the center of the nanostripe. When a spin-polarized current $j$ along the $-x$ axis with a density of 2 × 10¹¹ A m⁻² is on, the skyrmion moves along the $x$ axis with an initial velocity $v_x$ = 18.2 m s⁻¹ (Fig. 1b). However, the velocity $v_x$ decreases gradually to zero, the motion distance $G_x$ reaches a maximum value of 388 nm at time $t$ = 37 ns. A reverse skyrmion motion along the $-x$ axis is obtained in the time dual of 37.5-75.5 ns. Finally, the skyrmion stops at the distance $G_x$ = 340 nm even the current $j$ is still on. Once the current $j$ is off, the skyrmion moves along the $-x$ axis with an initial velocity $v_x$ = – 18.2 m s$^{-1}$, whose magnitude is equal to that at $t = 0$ ns. The relaxed velocity $v_x$ decreases when the guiding center of the skyrmion $G_x$ decreases. Finally, the skyrmion stops at the initial center of the nanostripe, i.e. $G_x = 0$ nm. The results suggest that the skyrmion prefers to stay at the center of the nanostripe. To deduce the main interaction which prevents the skyrmion from moving away from the center, we also obtain the motion dependence of free energy, as shown in Fig. 1c. The total free energy of skyrmion is not isotropic but increases as $G_x$ increases. The exchange interaction $\varepsilon_{ex}$, Zeeman $\varepsilon_{zeeman}$, and uniaxial anisotropy energies $\varepsilon_u$ all decrease as $G_x$ increases. In contrast, only the demagnetization energy $\varepsilon_{dem}$ increases as $G_x$ increases. Thus, the dipole-dipole interaction forces the skyrmion to stay at the central location of the nanostripe.

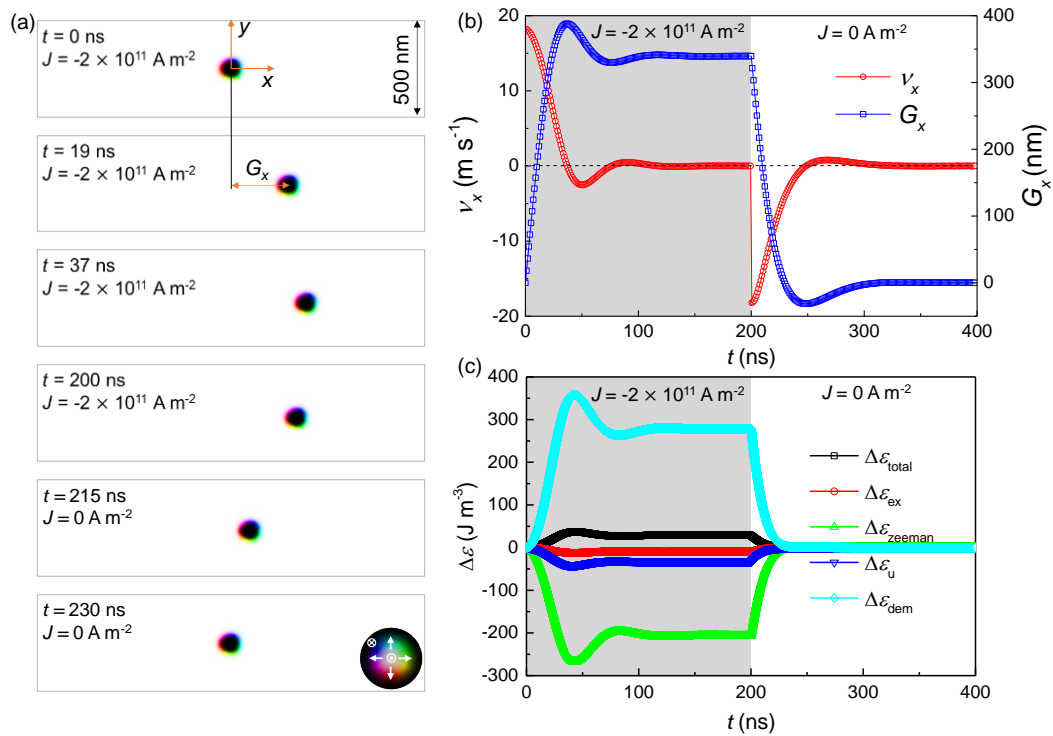

**FIG. 1.** (a) Snapshots for current-driven motions of a skyrmion in the middle layer ($z$ = 50 nm) of a ferromagnetic nanostripe. The central location of the nanostripe is set as the original zero-point. A spin-polarized current $j$ with a density of 2 × 10$^{11}$ A m$^{-2}$ is applied for 0 < $t$ < 200 ns and then is off for $t$ > 200 ns. The colors white and black represent out-of-plane up and down magnetization orientations, respectively, according to the colorwheel. (b) The skyrmion motion velocity along with the $x$ axis $v_x$ and the skyrmion guiding center $G_x$ as a function of time $t$. (c) The varied total free energy density $\Delta\varepsilon_{\text{total}}$, exchange energy density $\Delta\varepsilon_{\text{ex}}$, Zeeman energy density $\Delta\varepsilon_{\text{zeeman}}$, uniaxial anisotropy energy density $\Delta\varepsilon_{\text{u}}$, and demagnetization energy density $\Delta\varepsilon_{\text{dem}}$ as a function of time $t$ during the motion. The gray (0 < $t$ < 200 ns) and white ($t$ > 200 ns) regions in (b) and (c) imply that the current is on and off, respectively. Magnetic field $B_{ext}$ = 450 mT is applied along the $z$ axis.

We further understand the anisotropic demagnetization energy by exploring three-dimensional (3D) spin distribution in the nanostripe, as shown in Fig. 2a. In ultrathin two-dimensional (2D) uniaxial ferromagnets, skyrmion bubbles are stabilized only for quality factor $\eta$ > 1.[23,29] $\eta$ is defined as the ratio of the uniaxial perpendicular anisotropy $K_u$ to easy-plane magnetic shape anisotropy of 2D magnets, i.e. $\eta = K_u/(\frac{1}{2}\mu_0 M_s^2)$. $\eta$ > 1 suggests that the perpendicular magnetic anisotropy $K_u$ wins over the easy-plane magnetic anisotropy, resulting in perpendicularly magnetized 2D magnets and the formation of skyrmion bubbles. As the dipole-induced easy-plane magnetic shape anisotropy decreases as the thickness of the magnet increases, skyrmion bubbles can thus be stabilized in 3D magnets even for $\eta$ < 1.[22] We noted that $\eta$ is 0.22 in Fig. 1. A previous study has shown that the magnetic ground state in ferromagnetic nanodisks with $\eta$ < 1 can be manipulated by thickness.[37] In our case of $\eta$ = 0.22, the skyrmion bubble cannot be stabilized when the thickness of the nanostripe is smaller than 65 nm for $\eta$ = 0.22. The small $\eta$ suggests that dipole-dipole interaction plays a vital role in stabilizing the skyrmion bubble. Despite that uniform ferromagnetic background magnetization is observed in the middle layer of the nanostripe (Figs. 1a and 2d), the background magnetizations in surface layers are non-uniformed (Figs. 2c and 2e), which is owing to the formation of cross-sectional closure spin distributions (Fig. 2b) to minimize demagnetization energy. The non-uniform background magnetizations suggest that the demagnetization energy is dependent on the skyrmion location $G_x$. We obtained the skyrmion location dependence of demagnetization energy by pinning the skyrmion at each location $G_x$ where the central spins of the skyrmion are frozen. Without the pinning, the skyrmion cannot be stabilized away from the center. As shown in Fig. 2f, the skyrmion would stay at $G_x$ = 0 nm where the demagnetization energy is minimum. The demagnetization energy increases as the skyrmion moves away from the center $G_x$ = 0 nm.

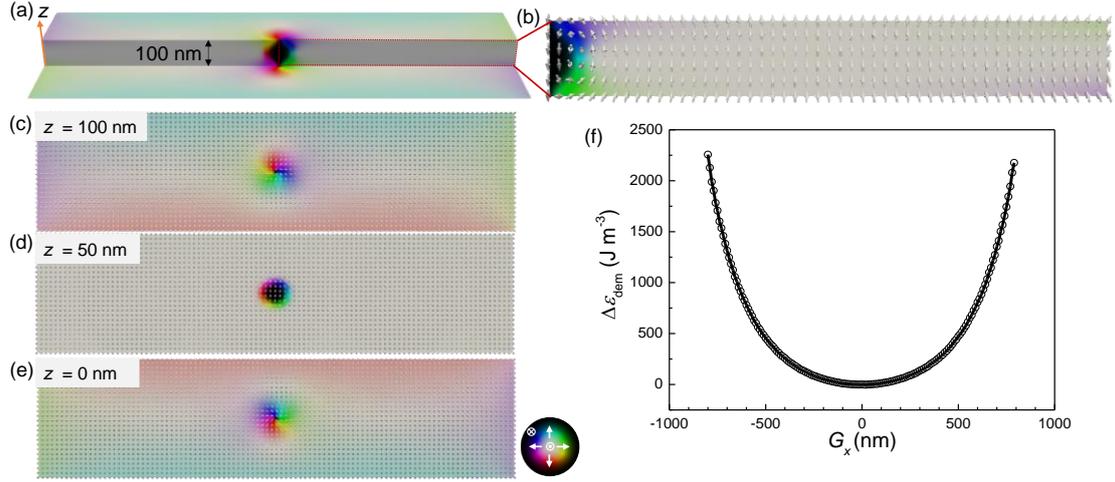

**FIG. 2.** (a) Three-dimensional spin textures of the skyrmion in the nanostripe. (b) Enlarged cross-sectional spin textures in the red zone of (a). Spin configurations at the top surface $z = 100$ nm (c), middle layer $z = 50$ nm (d), and bottom surface $z = 0$ nm (e). (f) Skyrmion location $G_x$ dependence of increased demagnetization energy $\Delta\varepsilon_{\text{dem}}$. Magnetic field $B_{ext} = 450$ mT is applied along the $z$ axis.

Figure 3 shows the current density dependence of dynamic motions of the skyrmion bubble in the nanostripe. Skyrmions are particle-like topological magnetic objects, whose motions are described using Thiele's collective coordinate approach.[4,17,38,39] Without the consideration of the demagnetization field, the velocity of the skyrmion bubble in our simulated case can be expressed by[4]:

$$v_x = -\frac{g\mu_B}{2eM_s}j \qquad (2)$$

Here, the parameters $g$, $\mu_B$, and $e$ are the Landé factor, Bohr magneton, and electron charge, respectively. For the skyrmion is located at the center, the skyrmion is free in motion. Thus, the initial velocity $v_{x0}$ at $G_x = 0$ nm can be well fitted using the equation (1), as shown in Fig. 3b. When the skyrmion moves away from the center of the nanostripe, the repulsive force from the sample edge induced by the demagnetization field prevents the motion of the skyrmion driven by the current. Thus, the velocity decreases gradually to zero when the driving force is balanced with the demagnetization field. The balanced motion distance $G_{x\text{-}m}$ increases nonlinearly with the current density (Fig. 3b) as the increased demagnetization energy is not linear with the distance $G_x$ (Fig. 2f). The repulsive force from the border of geometry has been also demonstrated in chiral skyrmion motion in non-centrosymmetric magnets[40,41]. But the repulsive force plays a role only when the chiral skyrmion gets very close to the edge. In contrast, for the dipolar-stabilized skyrmion bubbles in our case, the repulsive force works even if the skyrmion is 1000 nm away from the border. Noted that the skyrmion shrinks when getting approaching the edge of the nanostripe because of the repulsive force from the edge[17], as shown inset of Fig. 3a.

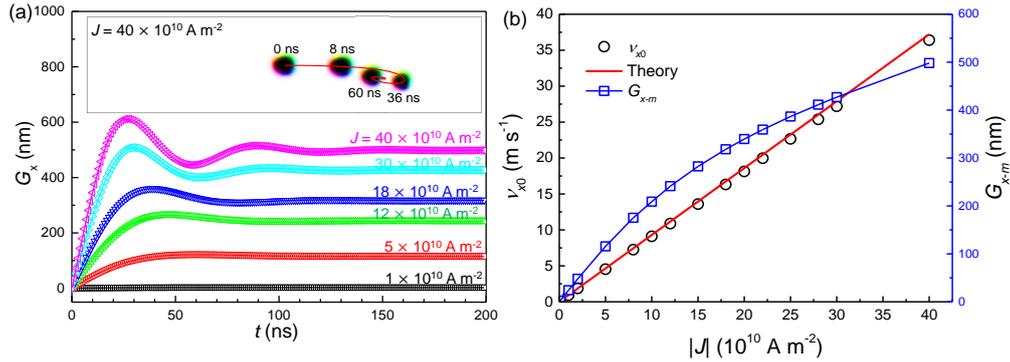

**FIG. 3.** (a) Skyrmion locations $x$ as a function of time $t$ for current density varying from $1 \times 10^{10}$ to $40 \times 10^{10}$ A m$^{-2}$. Inset of (a) shows the motion trajectory of the skyrmion bubble driven by the current with a density of $40 \times 10^{10}$ A m$^{-2}$. (b) Initial transient velocity at $t = 0$ ns $v_{x0}$ and dynamical equilibrium distance $G_{x\text{-}m}$ as a function

of current density. The red curve shows the fitting velocity based on equation (2).

Magnetic field $B_{ext}$ = 450 mT is applied along the $z$ axis.

The current-induced dynamic motions of the skyrmion bubble in the ferromagnetic nanostripe with $\eta$ = 0.22 could be applied for artificial spiking neurons.[33-36] Artificial neural networks are promising in dealing with massive-parallel tasks including pattern recognition and big data mining.[42] Artificial neural networks typically comprise synapses for memory and neurons for computing (Fig. 4a). Synapses have been experimentally demonstrated by current-controlled skyrmion counts.[36] The implementation of artificial neurons is mostly realized by integrating many transistors on Si-based circuits with a high energy cost and low device density.[43] The high mobility and small size of skyrmions enable promising high-efficient skyrmion-baed neuron devices.[33,34] The leakage-like functions of skyrmion-baed neurons are proposed by the artificial design of energy gradients, which are usually obtained by very complex geometrical designs including wedge-shaped nanotracks and gradient anisotropic films.[33,34] We will demonstrate the realization of the leakage-integrate-fire (LIF) behavior of neurons (Figs. 4b-4d ) using a dipolar skyrmion in the simple regular nanostripe.[44] Figure 4e shows the schematic neuron based on the achiral skyrmion bubble. We used pulsed spin-polarized currents as the input of spiking neuron $I_{\text{Spk-in}}$. The membrane potential $V_{\text{mem}}$ is represented by the driving distance $G_x$ of the skyrmion. Once the skyrmion enters the 100-nm width detection region whose center is 500 nm away from the center of the nanostripe, the spiking neuron fires an output $V_{out}$ that is represented by the measured topological Hall voltage $V_{THE}$.[45-49]

We considered period and stochastic waveforms of the input current stimuli (Fig. 4f), the distance $G_x$ of the skyrmion integrates when the current is on and leakages when the current is off (Fig. 4g). Once $G_x$ reaches 500 nm, the device fires an output topological Hall voltage (Fig. 4h) that is proportional to the total charge in the detection region:[45-49]

$$V_{THE} \propto \iiint_{x\,=\,450\,\text{nm}}^{x\,=\,550\,\text{nm}} q \, \mathrm{d}x\mathrm{d}y\mathrm{d}z \quad (3)$$

The device is then reset with zero current, the skyrmion will finally move back to the initial central location $G_x = 0$ nm because of the repulsive force of edge induced by the demagnetization field. Thus, the LIF behavior of artificial spiking neurons can well be mimicked by the dynamic motion of dipolar skyrmion driven by current stimuli. The LIF-like behavior is mainly contributed by the anisotropic demagnetization energy (Fig. 2f) in the confined nanostripe. When the periodical boundary condition along the $x$ axis is applied in our simulation, the demagnetization energy is then independent of the skyrmion bubble location. Thus, the skyrmion bubble can be steady driven by the current without the LIF-like behavior, as shown in Fig. 5a. Furthermore, because the anisotropic demagnetization energy as a function of skyrmion location remains when considering the skyrmion Hall effects and thicker thickness, the LIF-like behavior in $\eta < 1$ cases can be also achieved for $\alpha \neq \beta$ (Fig.

5b) and in thicker nanostripe (Fig. 5c).

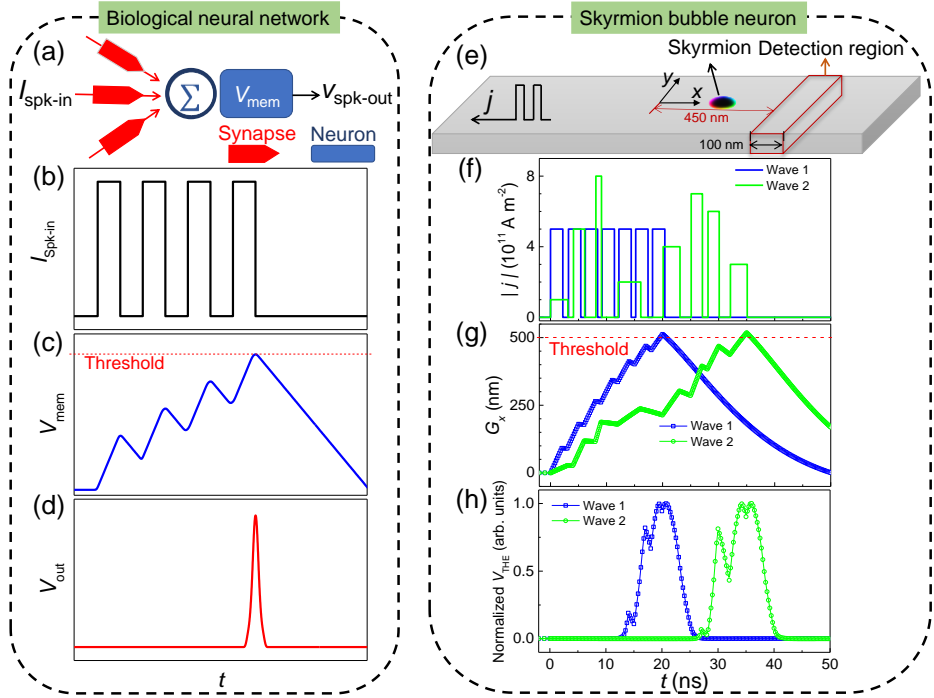

**FIG. 4.** (a) Schematic biological neural network includes neurons for memory and synapses for computing. (b)-(d) LIF behavior of artificial spiking neurons. (e) Schematic neuron based on current-induced skyrmion bubble motion in a nanostripe. The red box indicates the detection zone based on the topological Hall effect. (f) Period (blue curve) and stochastic (green curve) current stimuli are used as the input of spiking neurons $I_{\text{Spk-in}}$. (g) Motions of the skyrmion are driven by the period (blue curve) and stochastic (green curve) current stimuli. Once the location $G_x$ reaches 500 nm, the device is reset with $j = 0$ A m$^{-2}$ and fires an output topological Hall voltage $V_{\text{THE}}$. (h) The measured topological Hall voltage $V_{\text{THE}}$ in the detection region during the whole LIF behavior. Magnetic field $B_{ext} = 450$ mT is applied along the $z$ axis.

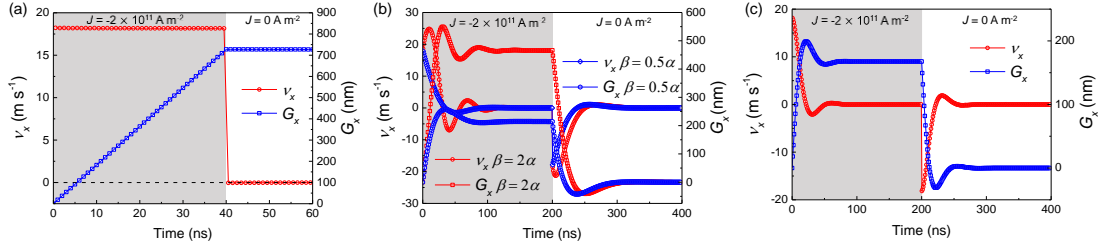

**FIG. 5.** (a) Current-driven dynamics of the skyrmion bubble in the nanostripe with the periodical boundary condition along the x axis. $\beta = \alpha$, thickness = 100 nm, and $B_{ext}$ = 450 mT. (b) Current-driven dynamics of the skyrmion bubble in the nanostripe for $\beta = 2\alpha$ and $\beta = 0.5\alpha$. Thickness = 100 nm, $B_{ext}$ = 450 mT without the periodical boundary condition. (c) Current-driven dynamics of the skyrmion bubble in the nanostripe with a thickness of 200 nm. $\beta = \alpha$, $B_{ext}$ = 450 mT without the periodical boundary condition. The gray and white regions imply that the current is on and off, respectively.

We have shown that the demagnetization field forces the skyrmion to stay at the central nanostripe, which could be used for neuron devices but is not applicable for racetrack memory.[11] We then discuss the current-induced coherent motion of the skyrmion in a stronger perpendicularly magnetized nanostripe. The quality factor is then modified by varying only the magnetic anisotropy $K_u$. By increasing the quality factor of the uniaxial ferromagnet, we observe that the background magnetizations turn to be uniform (Figs. 6a-6f). A skyrmion is initialed at $G_x$ = 500 nm, we then obtain the average relaxed velocity $v_{relax}$ in the first 10 ns after releasing the skyrmion. The relaxed velocity is proportional to the gradient of demagnetization energy along the $x$ axis, i.e. $v_{relax} \propto \partial \varepsilon_{dem}/\partial x$.[50] When the background magnetization turns to be uniform for high $\eta$ (Fig. 6c), the gradient of demagnetization energy decreases, thus the relaxed velocity $v_{relax}$ decreases as the quality factor increases and turns to zero for $\eta > 3.8$ (Fig. 6g), suggesting the repulsive force from the edge of geometry induced by the demagnetization field is strongly suppressed for increased $\eta$. We then test the current-induced coherent motion of a skyrmion for $\eta = 4.48$ (Fig. 6h). The skyrmion can be continuously driven by the current. Importantly, the skyrmion remains but does not move to the initial central location of the nanostripe when the current is off, providing the possible racetrack memory application.

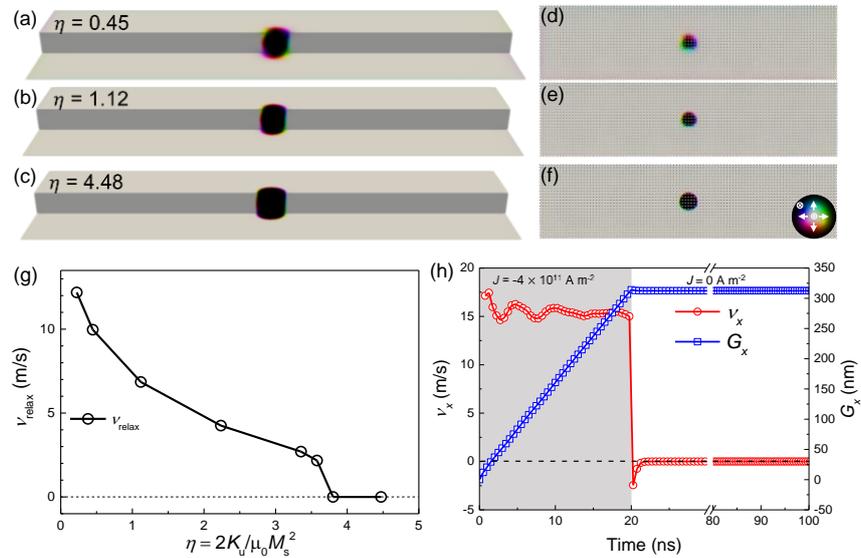

**FIG. 6.** 3D spin configurations for the skyrmion located at the center of the nanostripe for magnets with $\eta = 0.45$ at $B = 400$ mT (a), $\eta = 1.12$ at $B = 310$ mT (b), and $\eta = 4.48$ at 175 mT (c). The spin configurations at the top surface $z = 100$ nm for magnets with $\eta = 0.45$ (d), $\eta = 1.12$ (e), and $\eta = 4.48$ (f). (g) Quality factor dependence of initial averaged velocity in first 10 ns after releasing a skyrmion at $G_x = 500$ nm. (h)

Velocity and guiding center of a skyrmion driven by current in a magnet with $\eta = 4.48$ at $B = 175$ mT. The gray ($0 < t < 20$ ns) and white ($t > 20$ ns) regions in (h) imply that the current is on and off, respectively.

## 4. Conclusions

In summary, we have shown the role of dipole-dipole interaction on motion dynamics of skyrmion bubbles in achiral uniaxial ferromagnets. The LIF and coherent motion behaviors are demonstrated for uniaxial magnets with low and high quality factors, respectively. Besides, as the dipole-dipole interaction is also important in stabilizing antiskyrmion bubbles in MnPdPtSn and hybrid skyrmion bubbles in multi-layer films with perpendicular anisotropies,[51,52] our results should also provide a guide in their current-controlled dynamic manipulations.

**Acknowledgments**

This work was supported by the Natural Science Foundation of China, Grants No. 12174396 and 12104123.

**DATA AVAILABILITY**

The data that support the findings of this study are available from the corresponding author upon reasonable request.

**Conflict of interest**

The authors declare none.